# On the role of Grain Boundary Character in the Stress Corrosion Cracking of Nanoporous Gold Thin Films

*Aparna Saksena [*a], Ayman El-Zoka [b], Alaukik Saxena [a], Ezgi Hatipoglu [a], Jochen M. Schneider [c], Baptiste Gault [a, b]*

[a] Max-Planck-Institut für Eisenforschung GmbH, Max-Planck-Straβe 1, Düsseldorf 40237, Germany

[b] Department of Materials, Royal School of Mines, Imperial College London, Prince Consort Road, London, SW7 2BP, UK

[c] Materials Chemistry, RWTH Aachen University, Kopernikusstrasse. 10, Aachen 52074, Germany

*corresponding author: a.saksena@mpie.de

## Abstract

For its potential as a catalyst, nanoporous gold (NPG) prepared through dealloying of bulk Ag-Au alloys has been extensively investigated. NPG thin films can offer ease of handling, better tunability of the chemistry and microstructure of the nanoporous structure, and represent a more sustainable usage of scarce resources. These films are however prone to intergranular cracking during dealloying, limiting their stability and potential applications. Here, we set out to systematically investigate the grain boundaries in $Au_{28}Ag_{72}$ (± 2 at.%) thin films. We observe that a sample synthesized at 400 °C is at least 2.5 times less prone to cracking compared to a sample synthesized at room temperature. This correlates with a higher density of coincident site lattice grain boundaries, especially the density of coherent Σ3, increased, which appear resistant against cracking. Nanoscale compositional analysis of random high-angle grain boundaries reveals prominent Ag enrichment up to 77 at.%, whereas Σ3 coherent twin boundaries show Au enrichment of up to 30 at.%. The misorientation and the chemistry of grain boundaries hence affect their dealloying behavior, which in turn controls the cracking, and the possible longevity of NPG thin films for application in electrocatalysis.

Nanoporous materials have garnered a lot of interest for catalytic applications[1,2] as they provide a high specific surface area, enabling faster reaction kinetics[3]. Nanoporous gold (NPG), has been widely investigated for its applications in oxygen-assisted reactions such as alcohol coupling, methanol oxidation in fuel cells, and CO oxidation[4-6]. It is often synthesized by selective dissolution of Ag from an alloy of Ag-Au, through chemical or electrochemical corrosion called dealloying[7-9]. It is proposed that the selective dissolution of Ag in a completely miscible system such as Ag-Au is due to a spinodal decomposition at the surface–electrolyte interface[10]. Due to the surface diffusion of Au atoms[11], Ag is continuously exposed to the electrolyte leading to its dissolution and to the formation of a Au-rich, open-pore, bicontinuous network[10], provided that the concentration of Au is below a parting limit of 40 at.%[12-14]. The surface area can be altered by controlling the alloy composition and temperature of dealloying[15], making NPG an attractive alternative to tune the catalytic activity.

However, the cost of making bulk alloys, only for exploiting their surface, motivates limiting the weight of active material through the use of NPG thin films supported by a substrate[16]. The confined geometry adds a degree of freedom that allows for better control over the microstructure and local composition and allows for a more sustainable design of future catalysts [5,17]. In addition, microstructural defects, such as kinks and grain boundaries[18,19], can positively influence the catalytic activity. However, they are also the source of loss of stability during dealloying, and while there is interest in growing crack-free nanoporous films to improve their service durability[20-22], there is no clear understanding of the specific role of the numerous grain boundaries in limiting their viability.

NPG, especially NPG thin films, are susceptible to intergranular stress corrosion cracking[23-25]. Stress arises from the significant volume contraction (20–30%) during dealloying[20], with alloys leaner in noble metal cracking more[26]. For thin films, additional strain from the substrate makes them more prone to cracking[27]. Badwe et.al.[28] studied the separate role of corrosion and stress, and suggested that grain boundaries act as a fast diffusion pathway for dealloying to progress. They reported Au enrichment at grain boundaries far from the crack front. However, the degree of Au enrichment differs in the two studied grain boundaries. It is not clear whether this difference is a result of a difference in grain boundary misorientation angle and if the difference in the local composition results in differences in the cracking susceptibility. The composition of grain boundaries is expected to play a key role in their susceptibility to cracking and the enhanced catalytic activity. Additionally, for electrochemical dealloying, the critical potential required to initiate dealloying depends on the composition of the alloy[29]. Therefore, fluctuations in composition could influence the local dealloying kinetics.

Here, we aim to shed light on the influence of the grain boundary composition in the dealloying-induced cracking of Au-Ag thin films. We varied the synthesis temperature from room temperature to 400˚C, which changed the density of grain boundaries along with the distribution of misorientations. We reveal that crack propagation occurs preferentially at high angle grain boundaries segregated with Ag, whereas Σ3 grain boundaries are more resilient against cracking, which we relate to a slight segregation of Au. We rationalize this by inferences from the local composition variations at such defects.

Experimental methods

Ag-Au thin films were prepared by direct current magnetron sputtering in Ar (99.99% purity) at a pressure of 0.4 Pa, using the setup described in reference[30]. To improve adhesion a Ti layer was deposited onto Si (001) substrates by applying a power density of 2.5 W/cm$^2$. Thereafter Au$_{28}$Ag$_{72}$ films were deposited by co-sputtering from Ag (99.99% purity) and Au (99.9% purity) targets by applying a power density of 6.1 W/cm$^2$ and 2 W/cm$^2$ on Ag and Au targets, respectively. The base pressure was $\leq$ 7X10$^{-5}$ Pa and the substrates were rotated throughout the synthesis, to obtain a homogeneous composition and film thickness across the sample surface. Synthesis was performed at room temperature, i.e. with no intentional heating applied to the substrate, and at 400 °C. These films are referred to below as $AuAg^{RT}$ and $AuAg^{400}$, respectively.

Small sections (~7X10 mm$^2$) of the as-deposited thin films were cut and then chemically dealloyed by immersing them in concentrated nitric acid for 30 minutes. Subsequently, these samples were rinsed with distilled water to remove any residual acid and prevent further dealloying. Samples were then dried using nitrogen.

Microstructural characterization before and after dealloying was performed using a Zeiss Sigma scanning electron microscope (SEM, Carl Zeiss SMT, AG, Germany). Electron backscattered diffraction (EBSD) was obtained with a step size of 0.05 µm. The grain orientation and misorientation angle were analyzed by the TSL-OIM software package. A Zeiss Merlin SEM equipped with a Gemini-type field emission gun electron column was used for backscatter electron (BSE) imaging. The operating accelerated voltage was 15 kV with a probe current of 2–7.4 nA. Energy dispersive X-ray spectroscopy (EDX) was performed on Zeiss Sigma SEM with an acceleration voltage of 15 kV.

The local composition was characterized by atom probe tomography (APT) using a local electrode atom probe LEAP 5000 XR (Cameca Instrument Inc., Ametek, Madison, USA), in laser-pulsing mode, with a laser pulse energy of 60–70 pJ at a repetition rate of 100–125 kHz, and with an average detection rate of 5 ions per 1000 pulses. The specimens were maintained at 60 K during the measurement. The collected data was reconstructed and analyzed using AP Suite 6.3. APT specimens were prepared by following the protocol outlined in reference[31], using a dual-beam focused ion beam (FIB) FEI Helios 600i (Thermo Fisher Scientific Inc., Waltham, USA ) equipped with an EDAX Hikari Plus EBSD Camera. The final sharpening was guided by transmission Kikuchi diffraction (TKD) to verify the location and misorientation angle of the grain boundary in the APT specimen.

## Results

Figures 1 a and b show the cross-section of $AuAg^{RT}$ and $AuAg^{400}$ respectively, with the growth direction marked; the figures in the inset show the topography of the deposited films imaged via BSE. Both films are equiaxed instead of the characteristic columnar morphology. The grain size of $AuAg^{400}$ is more than three times larger, and grains appear more faceted, as readily visible from comparing the in the insets of Figure 1b with Figure 1 a.

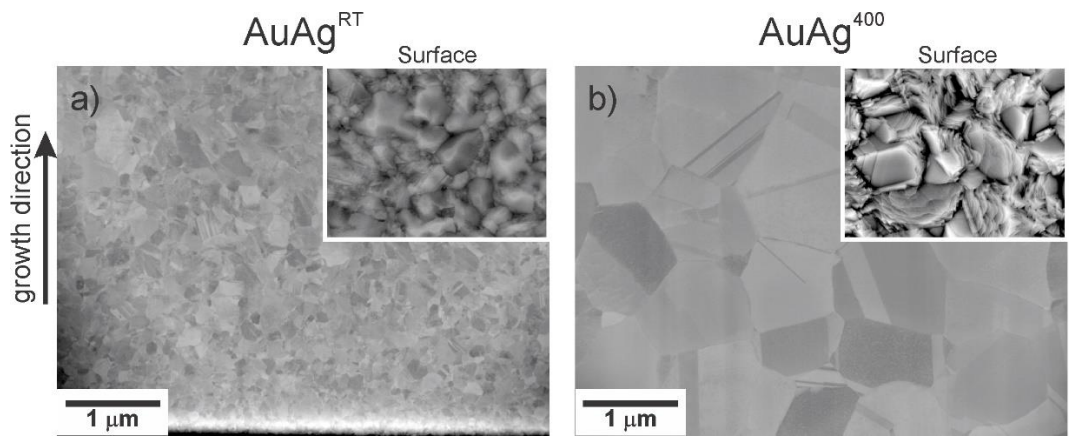

*Figure 1 shows the BSE image of the cross-section of $Au_{35}Ag_{65}$ deposited at room temperature (a)) and at 400 °C (b)). The inset figures show the top view (surface) of the deposited film at room temperature (inset-a)) and at 400 °C (inset-b)).*

The distribution of Au and Ag in the film was then studied by APT for $AuAg^{RT}$ and $AuAg^{400}$, as shown in Figures 2 a) and b) respectively. The (111) planes can be clearly visualized using the spatial distribution maps[32], and the interplanar spacing is calibrated to 0.235 nm, according to the theoretically expected (111) interplanar spacing. The concentration profile along the black dashed and solid arrows, is shown in Figure 2 c) and d) respectively. At thermodynamic equilibrium, Ag and Au are completely miscible at all temperatures and compositions[33], however, $AuAg^{RT}$ contains chemical modulations, particularly evident in the regions of the data with the highest spatial resolution i.e., at the crystallographic pole[34]. The wavelength of these chemical modulations ranges between 3.5–4 nm, while the amplitude of the modulation is ±8 at.% which agrees well with the deposition rate and the substrate rotation speed leading to deposition of alternate Au-rich layer and Ag-rich layer. This chemical modulation in $AuAg^{RT}$ contrasts with the homogeneous distribution of $AuAg^{400}$, Figure 2 d), and as assessed by frequency distribution analysis[35] (Figure S1). The average composition of $AuAg^{RT}$ differs slightly from that of $AuAg^{400}$, which can be attributed to chromatic aberrations[36] that occur in regions of high spatial resolution in APT, and is hence a measurement artefact. Figure S8 (supplementary information) shows that the mean composition of the film away from the crystallographic pole is similar to that of $AuAg^{400}$.

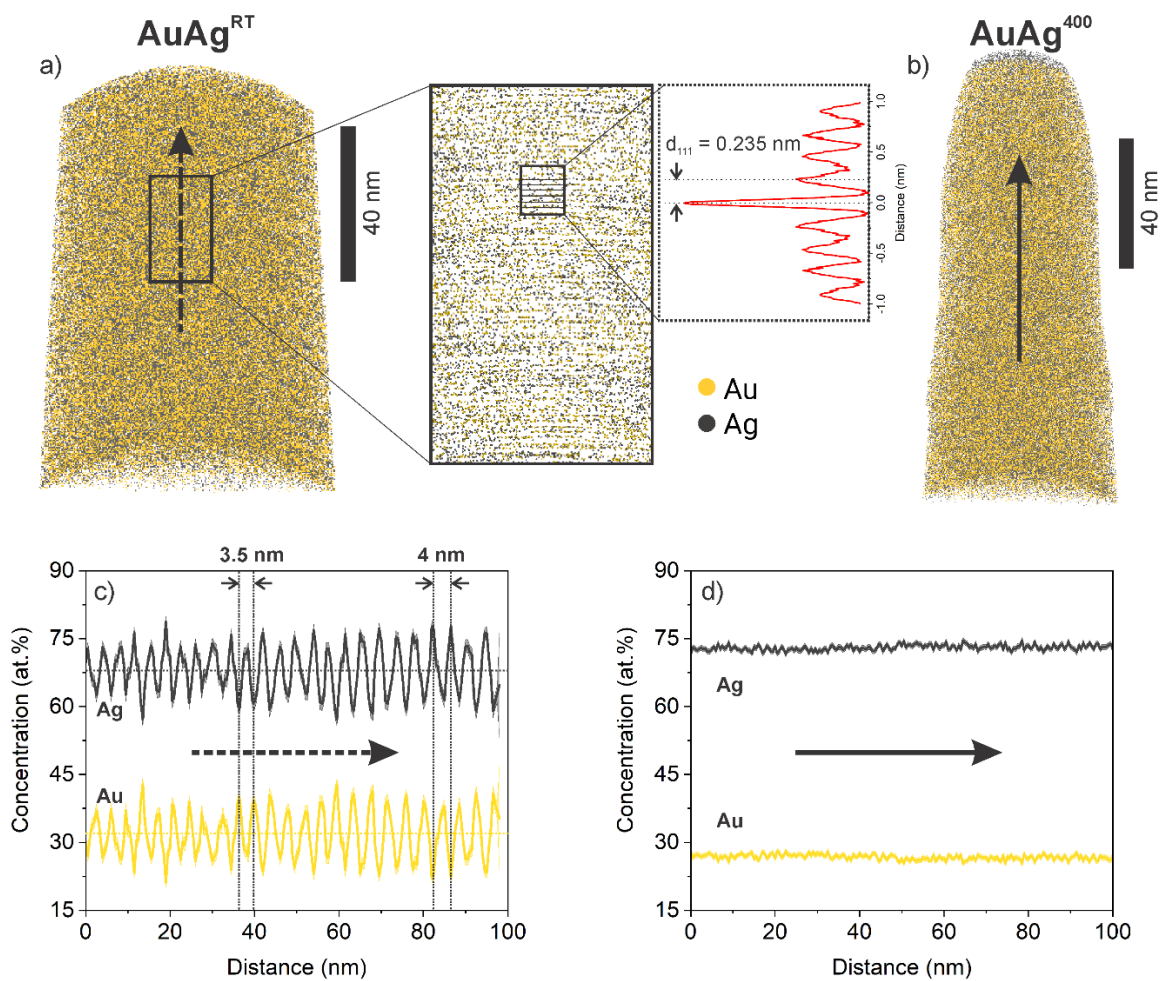

Figure 2 a) and b) shows the reconstructed datasets obtained via APT from $AuAg^{RT}$ and $AuAg^{400}$ respectively. c) and d) show the one-dimensional concentration profile across the region highlighted in a) and b) respectively.

The BSE image of $AuAg^{RT}$ and $AuAg^{400}$ after dealloying, displayed in Figure 3 a and b respectively, reveal a dense network of Au-rich nanoporous structure with a pore size of a few nanometers on average. The chemical modulation of Ag in $AuAg^{RT}$ remains within the minimum required composition (Au ≥ 40 at%) for dealloying percolation, which explains the formation of the nanoporous structure through chemical dealloying.

Cracks are also observed as dark features, such as those indicated by yellow arrows in Figure 3 a). The area fraction of cracks is 23–25% in $AuAg^{RT}$ and only 5–10% in $AuAg^{400}$, indicating a much higher crack density. The composition of the dealloyed films from EDX is 90 ± 2 at.% Au, with no apparent difference between the two films, suggesting that cracks do not significantly influence further dealloying at a scale that can be characterized by EDX. The larger grain size of $AuAg^{400}$ helps visualize that all cracks are intergranular, as shown in Figure 3 b). If some grain boundaries appear with no distinct contrast with the grain interior, marked by the black arrows; others exhibit a brighter contrast, as pointed by white arrows, and are devoid of cracks. This array of evidence points to a critical role of grain boundaries on corrosion-related cracking.

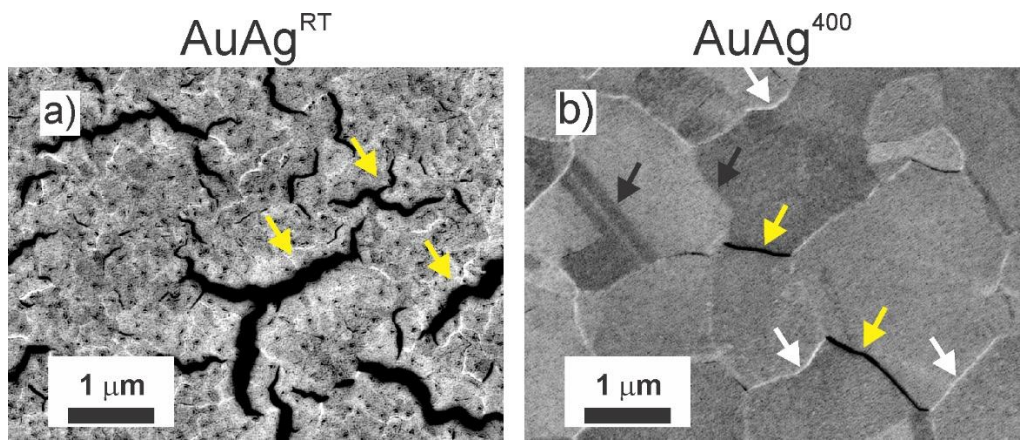

*Figure 3 a) and b) show the BSE image of $AuAg^{RT}$ and $AuAg^{400}$ respectively, after dealloying. The yellow and white arrows highlight the dark and bright contrast regions, respectively while the black arrows highlight the grain boundaries that show no distinct contrast.*

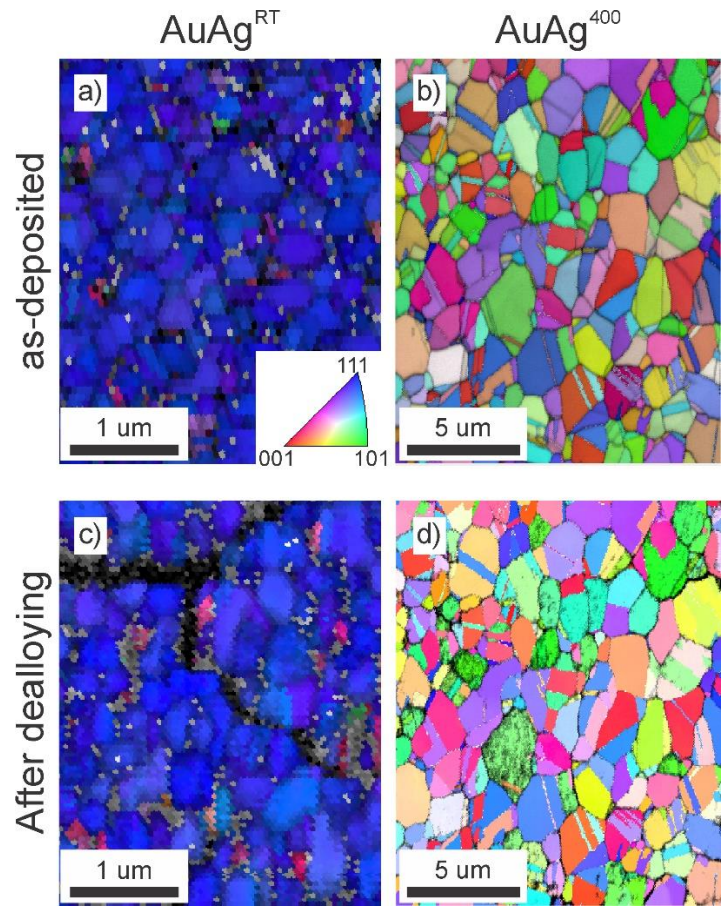

*Figure 4 a), and b) show the EBSD phase map and image quality map of the as-deposited $AuAg^{RT}$ and $AuAg^{400}$ respectively. c) and d) show the EBSD phase map after dealloying $AuAg^{RT}$ and $AuAg^{400}$ respectively. Grey pixels are due to the improper indexing of the EBS patterns caused by the small grain size. Note the difference in scale.*

Figures 4 a and b are the out-of-plane inverse pole figure map and image quality obtained from EBSD of the as-deposited $AuAg^{RT}$ and $AuAg^{400}$ respectively. Figure S2 shows the inverse pole figure for both in-plane and orthogonal axes of $AuAg^{RT}$, and the out-of-plane orientation is highly (111) textured while $AuAg^{400}$ is equiaxed. Figures 4 c and d are similar maps obtained after dealloying $AuAg^{RT}$ and $AuAg^{400}$ respectively. The darker regions correspond to the cracks seen in Figure 3. For $AuAg^{RT}$, a grain-to-grain comparison was not possible due to the small grain size, but the overall (111) texture persists after dealloying. For $AuAg^{400}$, the same region was imaged, evidencing that dealloyed grains have the orientation of the parent grain which is in agreement with the report from Jin *et.al*[37].

The total grain boundary length for $AuAg^{RT}$ is 2.5 times higher than $AuAg^{400}$. However, as estimated from Figure 3, the crack density is 2.5–4.5 times higher in the case of $AuAg^{RT}$. The number of coincident site lattice (CSL) grain boundaries, increased from 36% for $AuAg^{RT}$ to 60% for $AuAg^{400}$, with the fraction of ∑3 boundaries increasing from 70% for $AuAg^{RT}$ to 88% for $AuAg^{400}$. These ∑3 boundaries can often be indicative of coherent or incoherent twin boundaries, which is expected from the low stacking fault energy in the Ag-Au system[38]. We then performed image analysis based on pixel intensity thresholding, to seek a direct correlation between cracked grain boundaries in the BSE images after dealloying with the grain boundary misorientation angle obtained from EBSD. For the analyzed area, most of the cracks propagated through high-angle grain boundaries. Only 16% of the cracks, propagated through ∑3 grain boundaries, which constitute 55% of the total grain boundary length in $AuAg^{400}$.

CSL, and specifically ∑3 grain boundaries, exhibit high coherency and low energy, making them less susceptible to stress corrosion cracking. Beyond structural considerations, grain boundaries often do not have the same composition as the abutting grains[39-41]. We hence prepared APT specimens from random high-angle grain boundaries in $AuAg^{400}$ to reveal their local composition. Figure 5 a) shows the IPF and image quality obtained from TKD of an APT specimen containing a grain boundary with a 39° misorientation angle, and Figure 5 b) is the corresponding elemental map from the APT analysis.

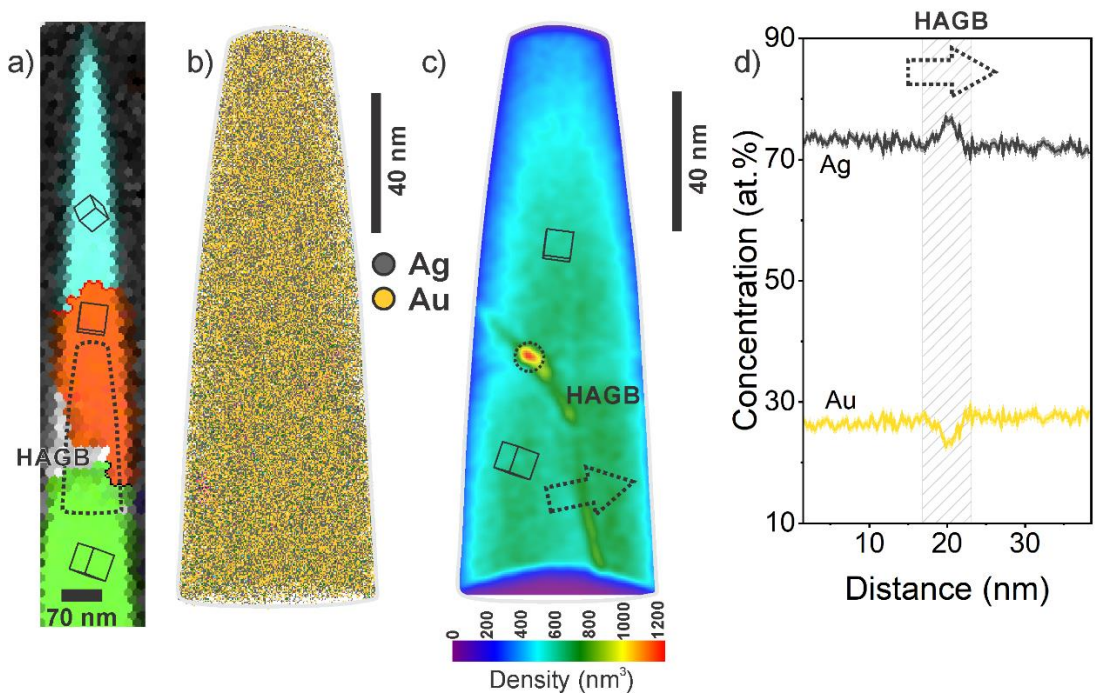

*Figure 5: a) IPF and image quality of the APT specimen, obtained via TKD. b) APT reconstruction of the measured specimen where the distribution of Au and Ag is shown. c) two-dimensional density map where the high-density region is identified. d) one-dimensional concentration profile along the region highlighted in c).*

A two-dimensional density map[42] is plotted where the grain boundary can be identified as the high-density region, as shown in Figure 5 c). A one-dimensional composition profile calculated across the grain boundary, as indicated by the arrow, reveals Ag enrichment of up to 77 at.% accompanied by a corresponding Au depletion. This Ag enrichment is 5% higher than the composition within the grain interior. Similar Ag enrichments are measured across another random high-angle grain boundaries (see Figure S4). The high-density region located along the grain boundary, marked by a dotted circle (also located in the grain for other measured specimens) is a nanoscale gold silicide (Supplementary Information) particle formed during the high-temperature synthesis[43]. Being in low quantity and primarily Au-rich, they are not expected to directly affect the dealloying behavior.

Figure 6 a) shows the IPF of an APT specimen containing a Σ3, coherent twin boundary (60° [111] 112), the local composition of which is measured via APT subsequently The distribution of Au and Ag in the corresponding APT reconstruction is displayed in Figure 6 b). Using a two-dimensional density map facilitated the identification of the twin boundary as a high-density region. A one-dimensional concentration profile calculated along the arrow is plotted in Figure 6 d). The concentration profile indicates a slight Au enrichment of up to 30.1 at.%. Compared to the composition of the grain itself, the Au enrichment is up to 3.5% higher. This is in stark contrast with the Ag enrichment observed at random high-angle grain boundaries. Au enrichment is observed at other Σ3 coherent twin boundary (Figure S5) however the region in the immediate vicinity of the twin boundary, also shows a slight Ag enrichment on both sides.

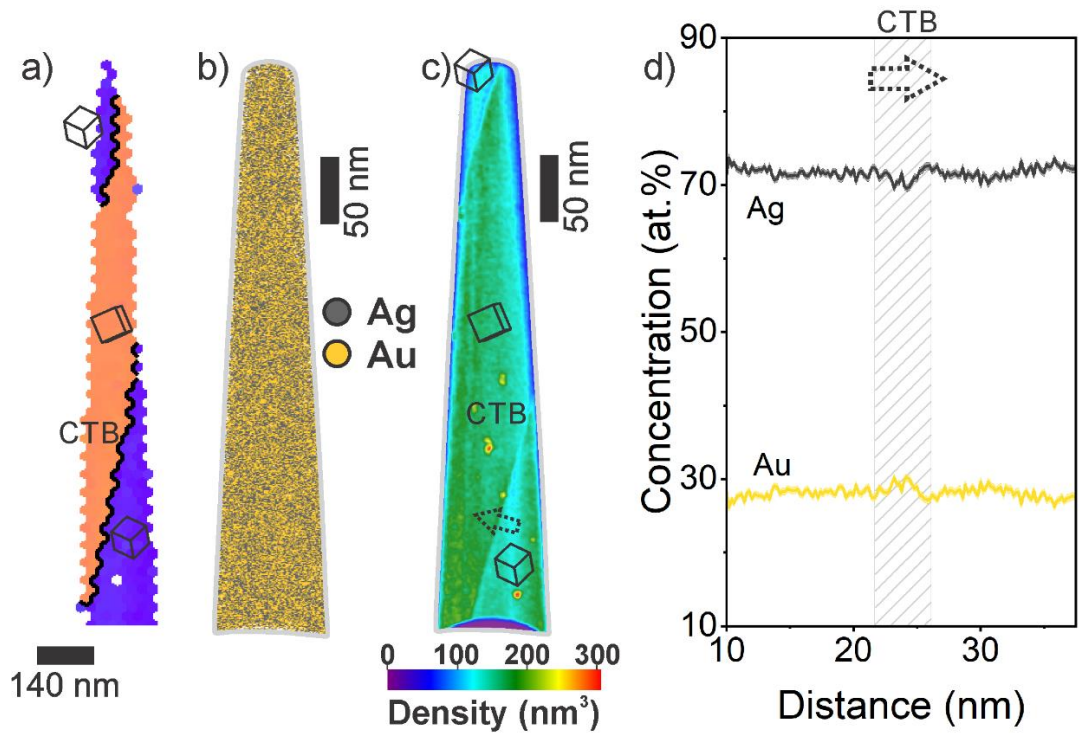

*Figure 6: a) IPF map of the APT specimen obtained after TKD. The grain boundary colored in black shows a Σ3 coherent twin boundary. b) APT reconstruction showing the distribution of Au and Ag. c) two-dimensional contour plot displaying the density variations. d) one-dimensional concentration profile across the higher density region highlighted in c).*

Segregation to Σ3 grain boundaries is known to be highly sensitive to the habit plane[39]. Figure 7 a) plots the IPF and image quality from TKD of an APT specimen containing two Σ3 incoherent twin boundaries, according to Brandon's criteria[44]. Figure 7 b) is the corresponding APT elemental map. Figure 7 c) is a two-dimensional point density map in which two twin boundaries marked as ICTB1 and ICTB2 can be observed, along with the three grains. The one-dimensional composition profile across ICTB1, i.e. along the black arrow in Figure 7 c), is plotted and shown in Figure 7 d). It shows a slight enrichment of Ag compared to the bulk, surrounded by a prominent Au enrichment, on both sides. The

Ag enrichment and the surrounding Au enrichment is up to 2.2% and 3.4% higher than the composition of the grain interior, respectively. A similar profile has been reported at twin boundaries in MnAl[45]. Figure S6 (supplementary information) highlights that even across a single twin boundary (ICTB1), the extent of enrichment differs locally in the range of 0 to 2.5 at.%. This proves once again that the chemistry of twin boundaries can be more complex than expected and varies depending on coherency and inclination angle[46]. For the present case, this can explain the different stress corrosion cracking susceptibility of Σ3 grain boundaries.

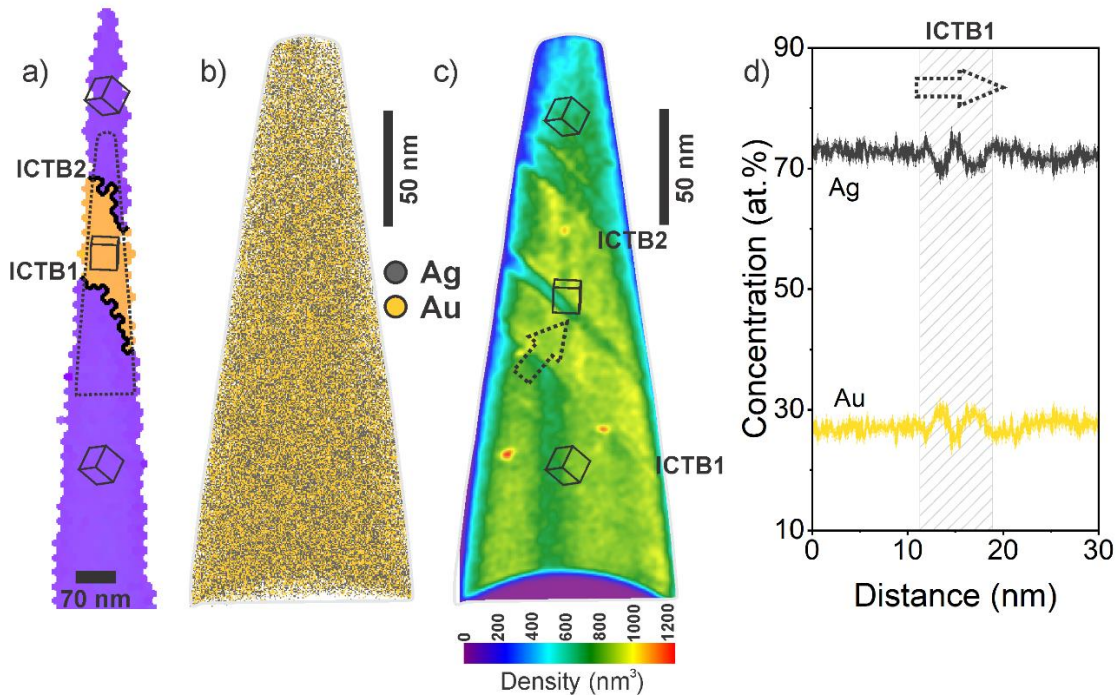

Figure 7 a) IPF map of the APT specimen obtained after TKD. The grain boundary marked in black shows a Σ3 incoherent twin boundary. b) APT reconstruction of the measured area highlighted in a), showing the distribution of Au and Ag. c) two-dimensional density map obtained from the reconstruction. d) one-dimensional concentration profile across the twin boundary highlighted in c).

## Discussion

*Thin film synthesis and microstructure*

Physical vapor deposition can lead to the formation of metastable phases. Room temperature growth, i.e. at 20% of melting point, is expected to be kinetically hindered. This explains the chemical modulations observed in $AuAg^{RT}$. However, $AuAg^{400}$ is expected to be closer to thermodynamic equilibrium. The synthesis temperature is 0.53 times the melting temperature which would imply bulk diffusion to be a dominant mechanism and therefore $AuAg^{400}$ is more homogeneous than $AuAg^{RT}$. This notion is also consistent with the increase in domain size as the growth temperature is increased. Additionally, the chemical enrichment at the grain boundaries in $AuAg^{400}$ should be closer to equilibrium and therefore should be representative of the bulk Au-Ag. Au enrichment at CSL boundaries is also in good agreement with the findings reported by Badwe et.al. on Σ5 boundaries[28].

*Segregation at the twin boundaries*

It has been reported that within the coherent twin boundaries, the elastic strain is distributed with alternating tensile and compression sites[47]. Where a larger atom, here Au, would preferentially segregate in the tensile sites, the smaller atom, here Ag would be preferred at the compression site, to minimize the elastic strain. However, this description is characteristic of the ideal, defect free grain boundaries. Recently, interfacial line defects accompanying the twin boundary called disconnections[48] have been reported, which are shown to be responsible for the change in segregation structure changes from a bilayer to a trilayer structure in Σ3 grain boundaries. This observation is in good agreement with the trilayer segregation we report here where the local chemistry across the twin boundaries shown in Figure 7 also shows a trilayer structure, segregation at the twin boundary plane and segregation at the region adjacent to the twin boundary plane.

*Influence of the local composition on the dealloying kinetics*

It has been reported that dealloying preferentially initiates at the grain boundaries and progresses subsequently in the grain interior[28]. Our findings indicate a prominent role of their chemistry. Σ3 boundaries show either Au enrichment or a very slight Ag enrichment surrounded by Au, which can explain their increased resistance to corrosion. The nanoporous network is also expected to be denser in the regions with higher Au concentration[26]. However, random high-angle grain boundaries show a significant Ag enrichment of up to 77 at.%, from a baseline at 72 at%. Therefore, it is reasonable to assume that the random high-angle grain boundaries will be subjected to faster dealloying kinetics and dissolution, and will hence be weaker against crack nucleation and propagation.

The critical potential for electrochemical dealloying depends strongly on the composition of the parent alloy[29], where a higher Au concentration requires a higher critical potential to initiate dealloying. This could make the controlled formation of NPG more challenging for these samples in the future. Considering the findings from Sieradzki *et.al.*[29], $Au_{28}Ag_{72}$ thin film would require a critical potential of at least 460 mV in a 1M $Ag^+$/Ag solution, to initiate dealloying. However, a 22% lower potential would already initiate Ag dissolution from the high angle grain boundaries while up to a 10% higher potential is required to initiate dealloying from the Σ3 grain boundaries. These insights on local chemical heterogeneities in the sample could lead to local variation in the dealloying kinetics which have not been considered at such a length scale.

## Conclusion

Here we investigated the role of grain boundaries in the stress corrosion cracking of Au-Ag thin films due to dealloying. $Au_{28}Ag_{72}$ (± 2 at.%) thin films were synthesized at room temperature and 400 °C, altering the grain size and subsequently grain boundary length and grain boundary character. The crack density was 2.5–4.5 times higher for the thin film synthesized at room temperature compared to the one synthesized at 400 °C. Where all the cracks were identified to be intergranular, they preferentially nucleated/ propagated through high angle grain boundaries. CSL boundaries especially Σ3 grain boundaries were observed to be more immune towards cracking. APT provided insights on the local chemistry of the high angle grain boundaries showing a significant Ag enrichment of up to 77 at.% whereas the Σ3 coherent twin boundaries showed up to 30 at.% of Au enrichment. These grain boundary character-dependent local variations in chemistry helped in rationalizing cracking mechanism which can thereby aid in designing mechanically stable NPG thin films for catalytic applications.

## Acknowledgement

The authors gratefully acknowledge Uwe Tezins, Andreas Sturm and Christian Bross for their support to the FIB and APT facilities at MPIE. ApS. acknowledges the financial support from Deutsche Forschungsgemeinschaft (DFG) under the collaborative research centre SFB/TR 103. AEZ and BG acknowledge the financial support from the German Research Foundation (DFG) through DIP Project No. 450800666. AIS appreciates funding by Helmholtz School for Data Science in Life, Earth and Energy (HDS-LEE). EH acknowledges support from IMPRS SusMet.